\documentclass[aps,prd,amsmath,amssymb,11pt]{revtex4}
\usepackage{graphicx}
\usepackage{tcolorbox}
\usepackage{bm}
\def\journal#1#2#3#4{{#1} {\bf #2}, #3 (#4)}
\newcommand{\be}{\begin{equation}}
\newcommand{\ee}{\end{equation}}
\newcommand{\bea}{\begin{eqnarray}}
\newcommand{\eea}{\end{eqnarray}}
\newcommand{\hf}{\frac12}
\newcommand{\nn}{\cr}
\def\eq#1{(\ref{#1})}
\def\fd#1#2{\frac{\delta#1}{\delta#2}}
\def\fdd#1#2#3{\frac{\delta^2#1}{\delta#2\delta#3}}
\def\pd#1#2{\frac{\partial#1}{\partial#2}}
\def\la{\langle}
\def\ra{\rangle}
\def\mr#1{{\mathrm{#1}}}
\def\ord#1{{\cal O}(#1)}

\def\dt{{\Delta t}}
\def\hj{{\hat j}}
\def\hx{\hat x}
\def\hy{\hat y}
\def\hD{{\hat D}}
\def\ih{\frac{i}{\hbar}}
\def\sign{\mr{sign}}

\def\Tr{{\mr{Tr}}}

\begin{document}
\title{Action for classical, quantum, closed and open systems}
\author{Janos Polonyi}
\affiliation{Strasbourg University, CNRS-IPHC,23 rue du Loess, BP28 67037 Strasbourg Cedex 2 France}
\date{\today}
\begin{abstract}
It is well known that the action functional can be used to define classical, quantum, closed, and open dynamics in a generalization of the variational principle and in the path integral formalism in classical and quantum dynamics, respectively. These schemes are based on an unusual feature, a formal redoubling of the degrees of freedom. Several arguments to motivate the redoubling are put forward in classical and quantum mechanics to demonstrate that such a formalism is natural.
\end{abstract}
\maketitle

\section{Introduction}
To understand the transition between the classical and the quantum physics one needs at least a common formalism, applicable in both domains. This is a wonderful problem because the classical limit of quantum system is driven by the interactions with a large environment, in other word the quantum system obeys open dynamics. Hence we need a CQCO formalism in mechanics which can handle Classical, Quantum, Closed and Open systems.

The search for simpler schemes, in particular for a CCO formalism covering classical closed and open systems, has started a century ago by attempting to describe the forces acting on electric charges by the help of the variational principle where the electromagnetic field was replaced by an effective non-local interaction \cite{schwarzschild,ritz,tetrode}. However an inconsistency arise because the electromagnetic forces obtained in such a manner are the sum of retarded and advanced contributions. The idea of the electromagnetic force resulting from a time reversal invariant action at a distance has been advanced further by renouncing the starting point, the variational principle, and by putting the burden on the absorber charges which were supposed to absorb completely the in-falling electromagnetic radiation \cite{wheeler}. Summarising in contemporary terms: The naive relativistic generalization of local forces without fields is doomed to a failure due to the lack of retardation \cite{currie,cannon,leutwyler}. The problem can be solved formally by the introduction of constraints \cite{kerner, pauri}. However the existence of electromagnetic radiation and our incapability of a full control of the electromagnetic field at the time of the observation of the charge dynamics requires the introduction of non-mechanical field degrees of freedom. The use of the variational principle to capture their contributions leads to the time reversal invariant near field interaction \cite{schwarzschild,ritz,tetrode}. To incorporate retardation one needs the far field component but the full representation of the dynamical field degrees of freedom without full experimental control renders the charge dynamics open and excludes the use of the traditional variational principle.

In another standard CQC formalism the action is used in the variational principle and in the path integral formalism in the classical and the quantum case, respectively. However we can not follow the transition from quantum to classical mechanics without uncontroled, ie. open, interaction channels. In fact, a necessary condition of the classical limit, the loss of coherence between macroscopically different pure states, can not be reached by the unitary time evolution in a closed dynamics. The traditional path integral formalism provides only the transition amplitude between pure states in closed dynamics. This does not mean that we always need an explicit environment for classical limit: A closed, large many-body system can easily provide the environment needed for the classical behaviour of some of its collective modes. But the interactions generating a classical dynamics are open within the simpler effective dynamics of the collective modes in question.

The extension of the canonical formalism over open systems represents a challenge both on classical and quantum levels and is based on the assumption that the observed system and its environment together obey a closed dynamics. In the deterministic world of classical mechanics one is tempted to retreat to probabilistic description like in kinetic theory. The open quantum theory is aimed by extracting the time dependence of the reduced density matrix of the observed sub-system by projection operators and the result is an integro-differential equation of motion \cite{feshbach,nakajima,zwanzig,mori}. The complexity of this equation restricts its application for Markovian weak coupling expansion \cite{redfield,gaspard,chruscinski,breuerk,preverzev,timm,kidon}. Another level of difficulties appears by recalling that the density matrix has to satisfy more stringent relations than the wave function of a pure state and the positivity can not be assured in the non-Markovian case \cite{barnett,shabani,maniscalco}. The usual solution of this problem is the phenomenological characterization of the most general Markovian master equation to produce physically acceptable density matrix \cite{lindblad}.

Another approach to effective quantum dynamics is the Closed Time Path (CTP) scheme \cite{schw,keldysh} and the resulting QCO formalism allows us to employ the standard perturbation expansion based on the physically appealing Feynman graphs \cite{hu,kamenev}. The linear response formalism \cite{kubo} corresponds to the leading order CTP perturbation expansion in the external sources. This method can be applied even in the projector operator formalism \cite{gu,breuerm}, as well. A similar idea leads to the description of quantum systems at finite temperaure \cite{umezawa}.

A distinguished feature of this formalism is a formal redoubling of the degrees of freedom. A classical system with coordinate $x$ is redoubled into two doublers, $x_+$ and $x_-$, which can be paired up into a single CTP doublet $x\to\hx=(x_+,x_-)$. The appearance of the CTP doublets is non-intuitive and renders the mathematics unusually involved and slowed down the spread of the applications of the formalism. But any practitioner of CTP can convince himself or herself that the complications of this scheme always represent true physical elements of the rich dynamics of open systems.

Our intuition arises from the macroscopic world which is supposed to be derived from the underlying and only formally known quantum dynamics. If the redoubling is indeed an inherent part of quantum physics then its trace should be visible in classical mechanics, too. The CTP formalism with redoubling has been introduced in classical Lagrangian \cite{envindta} and Hamiltonian \cite{galley} formalism. Hence we already have the elements of a CQCO scheme, the CTP formalism, where the dynamics is defined by the action functional. The goal of the present work is to identify the points motivating the redoubling in a CQCO formalism based on the action functional.

The efficiency of the CTP formalism in connecting the domains C, Q, C and O has already been attested by a large number of works. An incomplete list includes the derivation of an exact master equation for a harmonic quantum Brownian motion \cite{hubm}, the applications to driven open quantum systems \cite{sieberer}, to the semiclassical description of the decoherence \cite{dyndec}, the calculation of the entanglement entropy in quantum field theory \cite{boyanovsky}, of the Abraham-Lorentz force of electrodynamics \cite{ed}, of the 2PI effective action \cite{berges} and the construction of the quantum renormalization group for field theories \cite{openqft}. There are applications in gravity, such as the derivation of stochastic gravity \cite{husg}, the calculation of quantum effects to gravity \cite{elias}, of the thermal effects on Schwarzschild geometry \cite{campos}, and of the cosmological quantum damping \cite{broda}.

The above discussion of the action at a distance interaction of charges shows that the time arrow plays an important role in open dynamics hence we start in section \ref{auxtarrs} with a brief recapitulation of some notions and challenges of the direction of time. In section \ref{atas} the argument (C1) is presented by generalizing the traditional variation principle in such a manner that acausal auxiliary conditions can be replaced by causal initial conditions. The way the time arrow is encoded by the modified variation principle is demonstrated by the Green functions, introduced in section \ref{stpgreenfs}. The argument (C2) for redoubling is given in section \ref{opeclmechs} by the construction of the action for open systems. The emerging non-conservative forces, the generalization of the traditional holonomic forces, provide argument (C3) for redoubling. In section \ref{ancillas} the vision of the redoubling as an ancilla to preserve the Noether theorem in a non-conservative dynamics, argument (C4), is outlined together of the transformation of any local differential equation into canonical form, argument (C5). The arguments (Q1) and (Q2) of redoubling, presented in section \ref{clqds}, refer to closed quantum dynamics, the need of representing the time arrow and the way probabilities emerge in Gleason's theorem, respectively. The construction of the effective action of the path integral for the reduced density matrix in open quantum dynamics provides for argument (Q3). These arguments are spelled out first here, the traditional argument (Q0) of CTP, mentioned here only in passing, is to arrive at a perturbation expansion for observables in the Heisenberg representation and for the causal Green functions. The results are briefly summarised in section \ref{summs}. The calculation of the two point Green function for a harmonic oscillator is presented in appendix \ref{hogfa}.

\section{Classical equations of motion and their time arrows}\label{auxtarrs}
Physical laws have two components, an equation of motion, and some auxiliary conditions. The latter installs a time arrow for the former. These two components and their relations are briefly surveyed in this section.

\subsection{Auxiliary conditions}
An equation of motion alone is not sufficient to  make predictions because it contains time derivatives hence one has to impose some auxiliary conditions. These two components, the equations of motion and its auxiliary conditions, are strictly separated. The former comes from our theories and the latter is chosen by the experimentalists. Such a strong separation may explain that a slight inconsistency of the variational principle about causality remained unnoticed: On the one hand, we use non-causal auxiliary conditions in the classical variational method by fixing the initial and the final coordinates. On the other hand, the Euler-Lagrange equations are used with causal initial conditions in the applications. This discrepancy is acceptable as long as the auxiliary conditions are indeed independent of the equations of motions.

However, due to our human and technical limitations, we can never observe a genuinely closed system where the auxiliary conditions are fully under our control. When a subsystem of a closed dynamics is observed then the equations of motion of the observed system and the auxiliary conditions of its invisible environment are irrevocably mixed. In fact, let us denote the observed and the unobserved coordinates of a closed bipartite classical system by $x$ and $y$, respectively, and assume the equations of motion $\ddot x=F(x,y)$, $\ddot y=G(x,y)$ with the auxiliary conditions $y(t_a)=y_i$, and $\dot y(t_a)=v_i$. To find the effective equation of motion of the observed system first we solve the environment equation of motion for an arbitrary system trajectory $y=y[x;y_i,v_i]$ and next we insert the solution into the system equation of motion $\ddot x=F(x,y[x;y_i,v_i])$. The resulting effective equation displays an explicit dependence on the environment auxiliary conditions. If the open dynamics is to be derived from an action principle, the best scheme to describe a large system of particles, then that principle must be based on initial conditions.

\subsection{Time arrows}\label{tasf}
Several time arrows can be defined \cite{zeh,halliwell,houghton,albeverio} and they are usually classified according to the domains of physics where they appear: Time arrows are known in cosmology, quantum mechanics, a thermodynamics, and electrodynamics. It remains to be seen if these time arrows are independent of each other or they stem from a common origin. A time arrow can be informational or causal: The former points in the direction we loose information, e.g.. the auxiliary conditions degrade, examples being the quantum mechanical and the thermodynamical time arrows. The latter is directed from the cause to its effects, an example being the electrodynamical time arrow, and will be called simply ``time arrow'' below. Finally, the time arrow can be internal or external relative to the dynamics where it is observed: The latter is introduced by the auxiliary conditions and the former corresponds to equations of motions with broken time reversal symmetry.

By imposing the auxiliary conditions at a given time the generated external time arrow points away from that time, i.e. we find different time arrows before and after the imposition of the auxiliary conditions. Such a double, time-dependent time arrow is a characteristic feature of the solution of local equations of motion and has caused some complications in finding the origin of the thermodynamical time arrow since the second law of thermodynamics applies in either directions. To avoid such a pathological cases one employs causal auxiliary conditions, either initial or final.

One would think that the experimental determination of a time arrow is trivial but it is actually rather challenging. The reason is that the concept of cause is not defined in physics since it implies an external intervention into the physical world. For instance the Newton equation describes a correlation between the states of a particle at different times rather than referring to cause and consequence. The cause is usually replaced by the supposedly the free will of the physicist in selecting the initial conditions for the experiment. In fact, the choice of the initial conditions must be arbitrary in some extent to prove or to disprove an equation of motion experimentally.

After having granted the independence of the experimentalists from the observed system one can identify the causal time arrow by the help of a time dependent external source $j(t)$ coupled to the system in a finite time interval identified by some reference clock. The causal time arrow $\tau=\pm1$ relative to the reference time, is determined by the direction of the time the external intervention leads to changes in the state of the system.

The existence of an internal time arrow, irreversibility, can be observed by recording the motion and by checking wether the time reversed motion, seen by the replaying the recording backward, satisfies the same equation of motion. The orientation, usually defined by the direction of the stable relaxing motion, is used to define the internal time arrow.

\section{Action principle with time arrow}\label{atas}
The goal of the variational principle is the selection of the observed trajectory from a set of possible trajectories, called variational trajectory space. This space consists of trajectories which are at least twice differentiable and satisfy the desired auxiliary conditions to make the choice unique and well defined. The problematic feature of this scheme is that the variational trajectory space is defined by non-causal auxiliary conditions, by fixing the initial and the final coordinates, hence no time arrow can be introduced in this scheme. The generalization of the action principle to allow the dynamics to handle its time arrow is presented in this section for a single one dimensional dynamical degree of freedom characterized by the coordinate $x$ and the Lagrangian $L=m\dot x^2/2-U(x)$.

To keep track of the independent equations we discretize the time interval $t_i\le t\le t_f$ by introducing a small time step $\dt$ and represent the trajectories $x(t)$ as vectors $\vec x$ of dimension $N+2$ with components $x_n=x(t_n)$, where $t_n=t_i+n\dt$, $n=0,\ldots,N+1$, $\dt=(t_f-t_i)/(N+1)$. The action
\be\label{clact}
S(\vec x)=\dt\sum_{n=1}^{N+1}\left[\frac{m}2\left(\frac{x_n-x_{n-1}}\dt\right)^2-U(x_n)\right]
\ee
yields the variational equations
\be\label{vareqs}
0=\frac{\partial S(\vec x)}{\partial x_n}=\begin{cases}x_0-x_1&n=0,\cr
2x_n-x_{n+1}-x_{n-1}-\frac{\dt^2}mU'(x_n)&1\le n\le N,\cr
x_{N+1}-x_N-\frac{\dt^2}mU'(x_{N+1})&n=N+1.\end{cases}
\ee
It is easy to check that the evaluation of the potential energy at an intermediate point $x_n^{(\eta)}=(1-\eta)x_n+\eta x_{n-1}$ leads to $\ord{\dt^2}$ changes in the equation of motion at the end points and to a $\ord{\dt^3}$ correction at the intermediate points and brings no changes in the limit $\dt\to0$. Hence the particular placing of the potential energy within the time step $t_{n-1}<t<t<t_n$ in the discretized action \eq{clact} is irrelevant in the continuum limit $\dt\to0$.

The variational equation yields the correct continuum limit for internal points, $1\le n\le N$, however there is a problem at the end points because the velocity is vanishing there as $\dt\to0$ in the absence of kinetic energy contribution to the action before $t_i$ and after $t_f$. This is not a problem if $x_0$ and $x_N$ are provided by the auxiliary conditions. But in the case of initial conditions $x_0$ and $x_1$ are fixed by the initial coordinate and velocity and we must use the variational equation for $x_{N+1}$ which is incomplete. How can we complete it?

\begin{figure}
\caption{The motion is followed by the trajectory $\hx(t)$ in the generalized action principle from the initial to the finial time. A time reversal is performed at the latter and we follow the motion until it retakes its time reversed initial conditions.}\label{ctppathf}
\includegraphics[scale=.38]{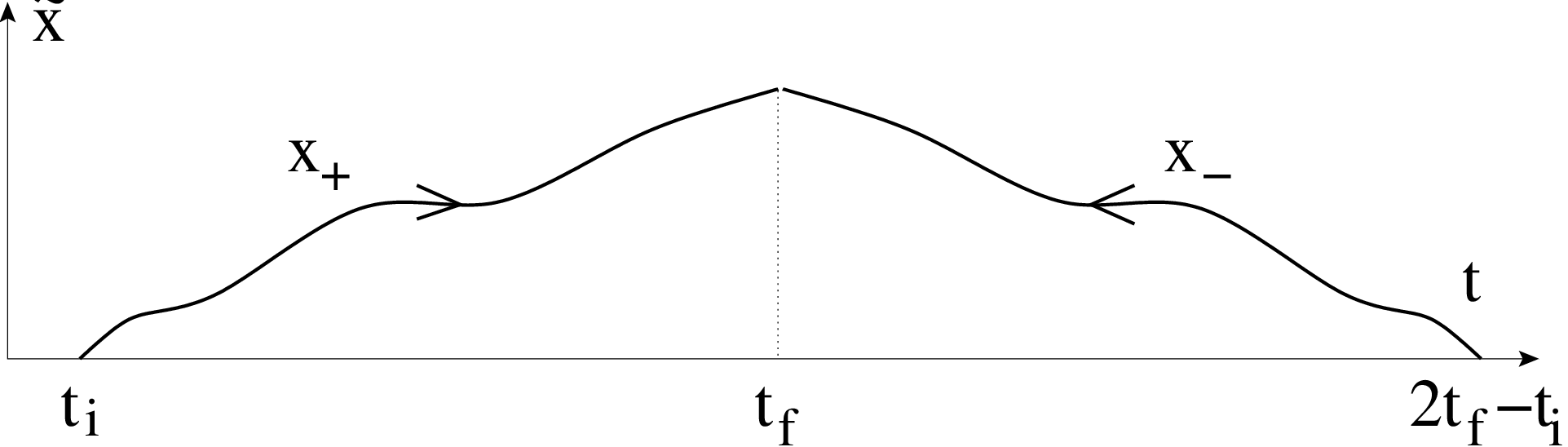}
\end{figure}

The solution is to make the last point an internal point with kinetic energy contribution from both neighbouring coordinate in time. Since $x_{N+1}$ is the last point of the motion it becomes an internal point only if there turn back in time. in other words, we make a time inversion $\dt\to-\dt$ and follow the motion backward in time. However the problem remains because there is still a last point in the backward moving part of the motion where the variational equation is incomplete. Notice that this is not the case anymore as soon as when we arrive back to the initial condition at $n=2N+2$. We can stop there because we already know the last two coordinate from the initial conditions, $x_{2N+2}=x_1$ and $x_{2N+3}=x_0$.

Let us check quickly the consistency: The number of the coordinates of the two trajectories is $2(N+2)$ which is reduced by the two initial conditions and the common end point to $2N-1$. Since we have $2N$ variational equation this seems to be an overdetermined problem. However the recursive solution of the equation of motion from the two end points, starting with $n=0,1$ and $n=2N+3,2N+2$ forward and backward in time, respectively, yields the same end points thus the number of independent equations is indeed $2N-1$.

Therefore the proposal is that we trace the trajectory twice: first forward in time then we make a time reversal and revisit the motion backward in time until we arrive back to the time reversed initial conditions as depicted in Fig. \ref{ctppathf}. Redoubling (C1) arises from breaking the trajectory of the roundtrip into two pieces $x(t')\to (x_+(t),x_-(t))$ with
\bea\label{roundt}
x_+(t)&=&x(t),\nn
x_-(t)&=&x(2t_f-t),
\eea
where $t_i\le t\le t_f$. The time flow backward in the second half of the roundtrip but the minus sign in front of the time variable in the second line of eqs. \eq{roundt} carries out another time inversion leading to the false impression that the time flows in the same direction in both trajectories. The action for the trajectory doublet $x_\pm(t)$ is
\be\label{cctpact}
S[\hx]=S[x_+]-S[x_-].
\ee
The variational trajectory space is defined by identical initial conditions for $x_+(t)$ and $x_-(t)$ and the final condition
\be\label{finalctp}
x_+(t_f)=x_-(t_f).
\ee
Note that the choice of the final time does not matter, the trajectory is $t_f$-independent for $t<t_f$.  The common final point justifies the name Closed Time Path of this scheme. The traditional action principle will be called Single Time Path (STP) formalism. The introduction of the CTP doublet $x\to\hx=(x_+,x_-)$ is {\em not} a redoubling of the physical degrees of freedom since we observe a single physical degree of freedom for twice and long time and even an irreversible the equation of motion sets $x_+(t)=x_-(t)$.

It is instructive to check the presence of a time arrow. Since the solution of the equation of motion makes the trajectories of the CTP copies identical an external source $j(t)=j_0\delta(t-t_0)$ with $t_i<t_0<t_f$ in the action induces a response for $t_0<t<2t_f-t_0$ on the trajectory $\hx(t)$ of Fig. \ref{ctppathf}, for $t_0<t<t_f$ in $\hx(t)$ and a unique causal time arrow is formed.

While the redoubling makes the use of the causal initial conditions possible in the variational principle it comes with a surprising high price. In fact, the redoubling of the coordinates seems to be out of proportion compared with the original problem, the change of the auxiliary conditions. But this is actually a reasonable price since the two trajectories satisfy the same equation of motion hence the effort to obtain them remains the same as in the traditional scheme.

It is instructive to see the generalization of other variational schemes, too. The Hamilton equations can be recovered as variational equations of the action
\be
S[x,p]=\int_{t_i}^{t_f}dt\left[p(t)\dot x(t)-H(p(t),x(t))\right]
\ee
where the auxiliary variable $p$ is introduced to arrive at first order equations of motion. As an example is the choice of $H=c\sqrt{m^2c^2+p^2}$ producing the dynamics of a free particle below the pair creation threshold. The action for the disrectized trajectories $x_n=x(t_n)$, $n=0,1,\ldots, N+1$, and $p_n=p(t_n)$, $n=1,\ldots, N+1$,
\be
S(\vec x,\vec p)=\dt\sum_{n=1}^{N+1}\left[p_n\frac{x_n-x_{n-1}}\dt-H(p_n,x_n)\right],
\ee
yields the variational equations
\bea
0&=&\frac{\partial S(\vec x,\vec p)}{\partial p_n}=\frac{x_n-x_{n-1}}\dt-\partial_pH(p_n,x_n),~~~1\le n=N+1\nn
0&=&\frac{\partial S(\vec x,\vec p)}{\partial x_n}=\begin{cases}-\frac{p_1}\dt&n=0,\cr
\frac{p_n-p_{n+1}}\dt-\partial_xH(p_n,x_n)&1\le n\le N,\cr
\frac{p_n}\dt-\partial_xH(p_n,x_n)&1\le n=N+1.\end{cases}
\eea

These equations have similar problems as eqs. \eq{vareqs} when initial conditions are used hence the remedy is similar, as well: The motion $t_i\le t\le t_f$ is followed for the time interval $t_i\le t'\le2t_f-t_i$ with a time reversal at $t'=t_f$ and the resulting trajectories $x(t')$ and $p(t')$ give rise to the CTP doublets \eq{roundt} and
\bea
p_+(t)&=&p(t),\nn
p_-(t)&=&-p(2t_f-t).
\eea
The CTP action is given by $S[\hx,\hat p]=S[x_+,p_+]-S[x_-,p_-]$ and the variational functional space is defined by the initial conditions $x_\pm(t_i)=x_i$, $p_\pm(t_i)=p_i$ and the closing of the doublet trajectories by the eq. \eq{finalctp} and $p_+(t_f)=p_-(t_f)$.

The Maupertuis variational principle is based on a variational functional space restricted by the energy conservation and has no immediate generalization for open system. However the related Fermat's principle \cite{landau} has mechanical application in finding the world line of a relativistic massive particle with a straightforward generalization for open dynamics \cite{cled}.

We close this section with a remark about locality corresponding to the time. The function of the coordinate at a given time, $f(x(t))$, and the function of the coordinate and the time derivatives of the trajectory up to a finite order, $f(t,x(t),dx(t)/dt,\ldots d^nx(t)/dt^n)$, are called ultra- and quasi-local, respectively. Non-local expressions involves either the the coordinates at different times or arbitrary high order time derivatives of the trajectory. An ultra-local Lagrangian yields algebraic variational equations hence the Lagrangians are quasi-local. The initial conditions of the Lagrangian formalism are quasi-local and those of the first order Hamilton equations are local. An importan virtue of the canonical formalism is that the phase space $(x,p)$ is the manifold of the ultra-local initial conditions imposed in such a manner that the phase-space trajectories, generated by the quasi-local equations of motion, do not intersect for short enough time.

\section{Green functions}\label{stpgreenfs}
A system of infinitely many Green function can be introduced for a classical dynamics and it offers two important advantages: It shows the role of the time arrow in a specially clear manner, and incorporates the initial conditions within the action. We continue to denote that coordinate of the observed degree of freedom by $x$ but its dynamics is defined by the STP action $S[x]$.

\subsection{STP Green functions}
We start with the traditional variation method by performing a functional Legendre transformation, $x(t)\to j(t)$, $S[x]\to W[j]$,
\be
W[j]=S[x]+\int_{t_i}^{t_f}dtx(t)j(t),
\ee
where the trajectory $x(t)$ is chosen by solving the equation of motion
\be
\fd{S[x]}{x(t)}=-j(t)
\ee
with fixed auxiliary conditions. The variational equation for $j$,
\be
\fd{W[j]}{j(t)}=x(t),
\ee
can be used to express $j$ in terms of $x$ and to construct the inverse functional Legendre transform
\be
S[x]=W[j]-\int_{t_i}^{t_f}dtx(t)j(t).
\ee
By restricting the book-keeping variable $j$ infinitesimal the functional $W[j]$ can be considered as formal functional power series,
\be
W[j]=\sum_{n=1}^\infty\frac1{n!}\int_{t_i}^{t_f}dt_1\cdots dt_nD_n(t_1,\ldots,t_n)j(t_1)\cdots j(t_n),
\ee
where the coefficient functions $D_n$ define the Green functions.

\begin{figure}
\centerline{\includegraphics[width=20pc]{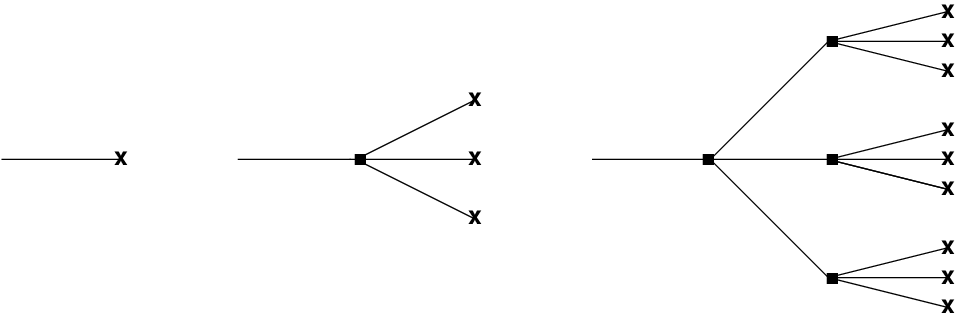}}
\caption{The iteration of the equation of motion \eq{intmeom} in terms of tree-graphs. The lines represent the Green function $D_2$, the dots stand for the vertex $g$, and the crosses denote the source, $j$.}\label{treegr}
\end{figure}

To find the physical roles of the Green functions let us consider the action
\be\label{weakint}
S[x]=\hf\int_{t_i}^{t_f}dtdt'x(t)K(t,t')x(t')-\frac{g}4\int_{t_i}^{t_f}dtx^4(t)+\int_{t_i}^{t_f}dtj(t)x(t)
\ee
where the kernel of the first integral is local in time with time translation invariance, $K(t,t')=K(d/dt)\delta(t-t')$, $K(z)$ being a second order polynomial. The kernel is taken symmetrical, $K(z)=K(-z)$, because the odd powers of the time derivative produce a boundary term in the action and drop out from the variational equations. Furthermore its roots are assumed to be imaginary, $K(\pm i\omega_0)=0$. The iterative solution of the equation of motion
\be\label{intmeom}
\int_{t_i}^{t_f}dtdt'K(t,t')x(t')=g\int_{t_i}^{t_f}dtx^3(t)-j(t)
\ee
which is reliable for a sufficiently short time is the infinite sum of tree graphs, the first three are shown in Fig. \ref{treegr}. Such a representation reveals that the Green function $D_n$ describes the $\ord{j^{n-1}}$ contribution to the trajectory.

\subsection{Mass-shell and off-shell modes}
To assess the role of the mass-shell modes we restrict our attention to a harmonic dynamics where $g=0$. The traditional formal $i\epsilon$ prescription to arrive at the causal Green functions is by the appropriately chosen infinitesimal imaginary shift of the poles of the frequency integral. This hides the origin of this problem, namely the need of a special treatment of a discrete spectrum point within the continuum.

The physical origin from the point of view of the variational problem is the absence of the solution of the homogeneous equation of motion in the action, $Kx_h=0$. The latter equation states that $x_h(t)$ belongs to the null space of the kernel $K$. The trajectories in the null-space will be called mass-shell modes as in field theory. While the mass-shell modes drop out from the action and thereby from the variational principle they are used in the traditional splitting of the solution into the sum of the general solution of the homogeneous equation of motion and a particular solution of the inhomogeneous case
\be\label{sephih}
x(t)=x_h(t)+x_{ih}(t).
\ee
In fact, the role of $x_h$ is to assure the desired auxiliary conditions. Hence the variational trajectory space consists of the solution of the inhomogeneous equation of motion satisfying the trivial Dirichlet condition, $x_{ih}(t_i)=x_{ih}(t_f)=0$, the off-shell modes.

The variational method proceeds within the variational trajectory space where $K$ is invertible and the near Green function $D^n(t,t')=D_2(t,t')$ is defined by the equations
\be\label{Ginverse}
\int_{t_i}^{t_f}dt'D^n(t_1,t')K(t',t_2)=\int_{t_i}^{t_f}dt'K(t_1,t')D^n(t',t_2)=\delta(t_1-t_2)
\ee
which holds up to on-shell corrections. Being the inverse of a symmetric operator $D^n$ is symmetric as well, $D^n(t,t')=D^n(t',t)$. The solution of the equation of motion of the harmonic model is given by
\be\label{harmsol}
x_{ih}(t)=-\int_{t_i}^{t_f}dt'D^n(t,t')j(t').
\ee

To recover translation symmetry in time we perform the limits $t_i\to-\infty$ and $t_f\to\infty$ where $D^n(t,t')=D^n(t-t')$ and
\be\label{ftr}
D^n(t)=\int_{-\infty}^\infty\frac{d\omega}{2\pi}e^{-it\omega}D^n(\omega).
\ee
Partial fraction decomposition can be used to bring the inverse of the kernel within the off-shell modes into the form
\be\label{ftrgfnt}
D^n(\omega)=\frac{Z}{\omega-\omega_0}-\frac{Z}{\omega+\omega_0}.
\ee
The frequency spectrum becomes continuous in this limit and the principal value prescription,
\be
D^n(\omega)=\frac{Z(\omega-\omega_0)}{(\omega-\omega_0)^2+\epsilon^2}
\ee
where the limit $\epsilon\to0$ is performed after the Fourier integral over the frequency, is applied to is skipped the null space resulting
\be\label{neargf}
D^n(t)=Z\sin\omega_0|t|.
\ee

The variational dynamics for the off-shell modes has non-oriented time due to the symmetry of the near Green function, $D^n(t,t')=D^n(t',t)$. The time arrow arises from the mass-shell modes in the decomposition \eq{sephih}, introduced ``by hand'', beyond the STP variational scheme. This decomposition is reflected in the definition of the retarded Green function $D^r=D^n+D^f$ where the far Green function $D^f$ acts within the null-space,
$D^f(t)=a\cos\omega_0t+b\sin\omega_0t$,
and the constants $a=0$ and $b=Z$ are determined by causality, $D^f(t)=\sign(t)D^n(t)$.

\subsection{CTP Green functions}
The variational space of the CTP formalism is defined by the initial conditions thus the time arrow is introduced in that space. The way this happens can clearly be seen by the help of the Green functions. As a preparation one introduces independent sources $\hj=(j_+,j_-)$ for the CTP copies, defines the functional Legendre transformation
\be\label{legendre}
W[\hj]=S[\hx]+\int_{t_i}^{t_f}dt\hx(t)\hj(t),
\ee
$\hx\hj=x_+j_++x_-j_-$, and $\hx(t)$ solves the equation of motion
\be\label{legendrev}
\fd{S[\hx]}{\hx(t)}=-\hj(t)
\ee
within the variational space, meaning identical initial conditions for $x_+$ and $x_-$ at $t_i$ and the final condition $x_+(t_f)=x_-(t_f)$.

The Green functions are defined by the functional Taylor series
\be\label{cgrfnct}
W[\hj]=\sum_{n=0}^\infty\frac1{n!}\int_{t_i}^{t_f}dt_1\cdots dt_nD_{n,\sigma_1,\ldots,\sigma_n}(t_1,\ldots,t_n)j_{\sigma_1}(t_1)\cdots j_{\sigma_n}(t_n).
\ee
The iterative solution of the equation of motion results the series of the tree graph of Fig. \ref{treegr} as for closed dynamics. The variational equation for $\hj$,
\be\label{ilegendrev}
\fd{W[\hj]}{\hj(t)}=\hx(t),
\ee
and its successive derivatives establish the same interpretation of the Green functions as in the case of closed dynamics. The two external sources $j_\pm$, are treated as independent in the functional Legendre transformation but by setting $j_+(t)=-j_-(t)$ after the variation leads to $x_+(t)=x_-(t)$.

A harmonic dynamics is defined by the action
\be\label{hoactcn}
S[\hx]=\hf\int_{t_i}^{t_f}dtdt'\hx(t)\hat K(t,t')\hx(t')+\int_{t_i}^{t_f}dt\hx(t)\hj(t)
\ee
and the solution of the equation of motion is
\be\label{ctpsol}
\hx(t)=-\int_{t_i}^{t_f}dt'\hD(t,t')\hj(t').
\ee
where $\hD=\hD_2$. The Green function is real at this point. To find identical copies of $x$ for physical sources, $j_\pm=\pm j$, one needs $D_{++}+D_{--}=D_{+-}+D_{-+}$. Since $D$ is symmetrical, $D_{\sigma_1,\sigma_2}(t_1,t_2)=D_{\sigma_2,\sigma_1}(t_2,t_1)$, the only solution of the latter equation is that both sides are vanishing,
\be\label{clgr}
\hD=\begin{pmatrix}D^n&-D^f\cr D^f&-D^n\end{pmatrix}
\ee
in terms of two real functions $D^n(t,t')=D^n(t',t)$ and $D^f(t,t')=-D^f(t',t)$. The solution \eq{ctpsol} for a physically realizable source is
\be
x(t)=-\int_{t_i}^{t_f}dt'D^r(t,t')j(t').
\ee
It is easy to see that the time arrow is properly installed in the iterative solution of the anharmonic model where the lines of Fig. \ref{treegr} stand for the retarded Green function.

\subsection{Generalized $\epsilon$-prescription}\label{epresrs}
The Green function defined for a finite time span $t_f-t_i<\infty$ is not translation invariant. To regain translation invariance in time one performs the limit $t_i\to-\infty$ and $t_f\to\infty$. However this limit creates another difficulty which is related to the null space of $\hat K$ where the Legendre transformation \eq{legendre} is ill-defined. While the generic discrete frequency spectrum $\omega_n=2\pi n/(t_f-t_i)$ for $t_f-t_i<\infty$ stays at finite distance away from the null space $\omega=\omega_0$ we have to make sure that  the action remains similarly non-degenerate as $t_f-t_i\to\infty$ . There are two different denegeneracies to avoid: A discrete degeneracy is in the null-space and it is lifted by the usual $\epsilon$-prescription, the introduction of an infinitesimal imaginary term in the STP action $S[x]\to S[x]\pm i\epsilon\int dtx^2(t)/2$. A continuous degeneracy of the CTP action $S[\hx]=S[x_+]-S[x_-]$ arises for $x_+(t)=x_-(t)$, it is lifted by using different signs for the imaginary part for the two doubler trajectories,
\be\label{cctpactf}
S[\hx]=S[x_+]-S[x_-]+i\frac\epsilon2\int_{t_i}^{t_f}dt[x_+^2(t)+x_-^2(t)].
\ee

The time reversal, a trivial reparametrization invariance of the motion, can be achieved in two steps: The time inversion $t\to-t$ of the trajectories and in the action is realized by the exchange of the two doublers, $(x_+,x_-)\to(x_-,x_+)$, and by the change of the sign $S\to-S$, respectively. Since the imaginary part of the action is identical for the two doublers we perform next a complex conjugation of the action $S\to S^*$. Therefore the symmetry under formal time reversal is expressed by the identity
\be\label{ctpsym}
S[x_+,x_-]=-S^*[x_-,x_+].
\ee

The properties of $\hD$ mentioned above for $t_f-t_i<0$ remains valid and now yield the block structure for the complex Green function
\be\label{cslgr}
\hD=\begin{pmatrix}D^n+iD^i&-D^f+iD^i\cr D^f+iD^i&-D^n+iD^i\end{pmatrix}
\ee
in terms of three real functions $D^n(t,t')=D^n(t',t)$, $D^i(t,t')=D^i(t',t)$, and $D^f(t,t')=-D^f(t',t)$. The time translation invariant Green function \eq{gfctinft} of a harmonic oscillator turns out to be
\be
\hD(\omega)=\frac1m\begin{pmatrix}\frac1{\omega^2-\omega_0^2+i\epsilon}&-i2\pi\Theta(-\omega_0)\delta(\omega^2-\omega_0^2)\cr-i2\pi\Theta(\omega_0)\delta(\omega^2-\omega_0^2)&-\frac1{\omega^2-\omega_0^2-i\epsilon}\end{pmatrix}
\ee
in the limit $t_i\to-\infty$ and $t_f\to\infty$. The variational trajectory space is defined by the generalized Dirichlet boundary conditions and the desired initial conditions at finite time can be achieved by an appropriate adiabatic turning on the external source.

The inverse Green function can be written in a similar form
\be
\hat K=\begin{pmatrix}K^n+iK^i&K^f-iK^i\cr-K^f-iK^i&-K^n+iK^i\end{pmatrix}
\ee
with
\bea
K^{\stackrel{r}{a}}&=&K^n\pm K^f=(D^{\stackrel{r}{a}})^{-1},\nn
K^i&=&-D^{r-1}D^iD^{a-1}.
\eea

The generalization of this procedure for a generic action STP $S[x]=S_0[x]+S_i[x]$ where $S_0$ corresponds to a harmonic oscillator and $S_i$ stands for the interaction produces the action
\be\label{genepspr}
S[\hx]=\hf\int_{-\infty}^\infty dt_1dt_2\hx(t_1)\hat K(t_1-t_2)\hx(t_2)+S_i[x_+]-S_i[x_-].
\ee
The imaginary part, the generalized $\epsilon$-prescription term,
\be
S_\epsilon[\hx]=\frac\epsilon\pi\int_{-\infty}^\infty dtdt'\frac{x^-(t)x^+(t')}{t-t'+i\epsilon}+\frac{i\epsilon}2\int_{-\infty}^\infty dt[x^{+2}(t)+x^{-2}(t)]
\ee
takes care of the CTP boundary conditions. The upshot is that the time translation symmetry breaking coupling \eq{finalctp} of the CTP doublet trajectories at the final time is spread over an infinitesimal time translation symmetrical coupling between the trajectories.

\section{Open classical mechanics}\label{opeclmechs}
The traditional variational principle was extended in closed dynamics to incorporate the time arrow. The next step is the generalization of the action for open dynamics.

\subsection{External time arrow}\label{inttarrws}
\begin{figure}
\caption{The time arrows of the system and the environment. The fat horizontal line indicates the time when the auxiliary conditions are imposed, and the corresponding time arrows are shown, as well. (a): $\tau_s=\tau_e=1$; (b): $\tau_s=\tau_e=-1$; (c): $\tau_s=1$, $\tau_e=-1$; (d): $\tau_s=1$, $\tau_e=-1$. The system coordinate is changed by a small amount at the time of the right oriented horizontal dashed line, the arrow representing the impact of the change on the environment. At another time, corresponding to the left oriented horizontal dashed line, the change of the environment trajectory feeds back to the system itself and appears as a self interaction within the system, induced by the environment.}\label{sectas}
\includegraphics[scale=.3]{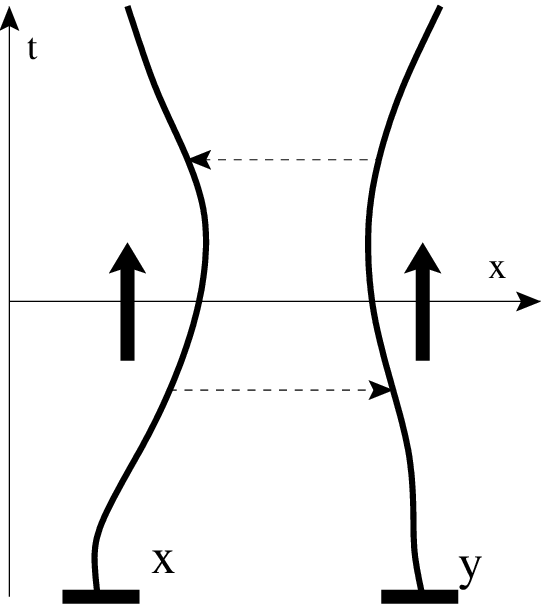}\hskip.5cm
\includegraphics[scale=.3]{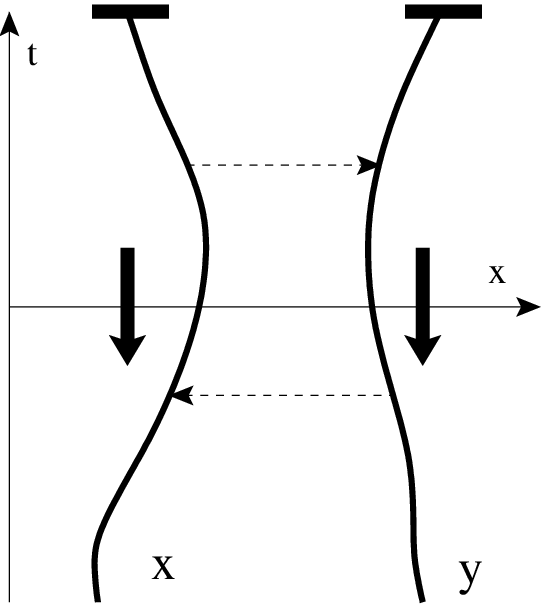}\hskip.5cm
\includegraphics[scale=.3]{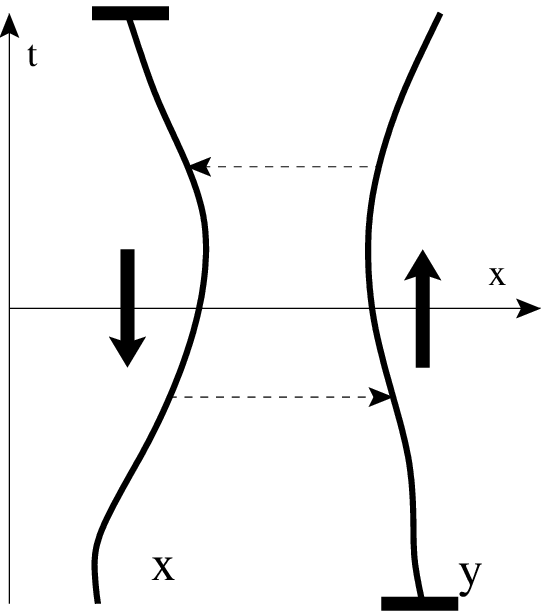}\hskip.5cm
\includegraphics[scale=.3]{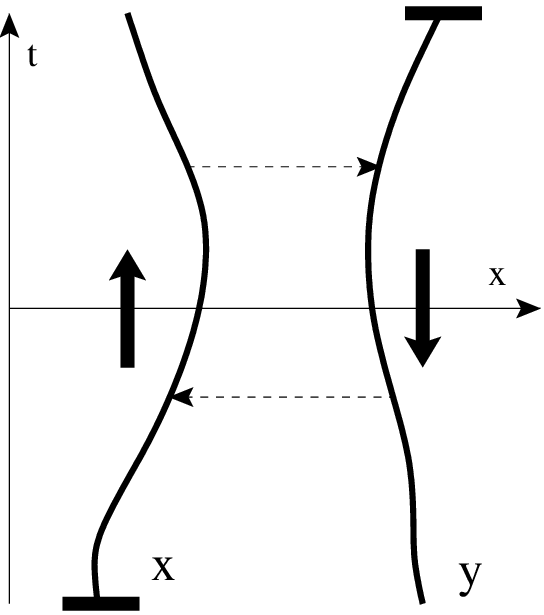}

(a)\hskip3.4cm(b)\hskip3.4cm(c)\hskip3.4cm(d)
\end{figure}

We start at the bipartite system mentioned in the Introduction. By imposing initial or final conditions in our mathematical modeling we can install a time arrow independently for the system and for the environment as long as they do not interact. Hence we can formally prepare parallel or anti-parallel flow of time for the non-interacting subsystems. The four possible orientation of the time are displayed in Fig. \ref{sectas} where the arrows indicate the system and the environment time arrows, $\tau_s$ and $\tau_e$, respectively in the absence of interactions.

When the system-environment interactions are turned on then the effective dynamics remains causal for $\tau_s=\tau_e$ but becomes acausal for $\tau_s=-\tau_e$. Expressed in another manner, the causal structure of the interactive system is the result of two time arrows, an external one set by the system auxiliary conditions the and an internal one inherited from the environment into the effective system dynamics. The effective dynamics is causal only for identical external and internal time arrows. Yet another way of summarising the lesson of Fig. \ref{sectas} is that a distinguished feature of open interaction channels is that they transfer an external time arrow from the environment into an internal one of the system. Hence the internal time arrow in a world with time reversal fundamental interactions is a signature of an environment.

The lesson is that the system-environment always breaks the time reversal invariance of an open dynamics. This actually follows from the proposed experimental detection of the time arrow mentioned in section \ref{tasf}, as well, since the recording does not show the environment. In fact, the time appears to run in opposite direction in the observed system and in the environment in the inverted replaying unlike to the original physical case.

\subsection{Semi-holonomic forces}
The conservative holonomic forces of the action principle are represented by a potential $U(x,\dot x)$ and are of the form
\be\label{hol}
F(x,\dot x)=-\partial_xU(x,\dot x)+\frac{d}{dt}\partial_{\dot x}U(x,\dot x).
\ee
To find their generalization, the open semi-holonomic forces, we start with a full closed dynamics of the observed system and its environment characterized by the STP action $S[x,y]$. The elimination of the environment is achieved by solving the equation of motion $\delta S[x,y]/\delta y(t)=0$ together with the environment initial conditions imposed at $t_i$ for a general system trajectory $x(t)$. The effective system action is obtained by inserting the solution, $y[x]$, into the action, $S_{eff}[x]=S[x,y[x]]$. The so far unspecified system trajectory is now selected by solving the variational equation of $S_{eff}[x]$ and the system auxiliary conditions. The effective equation of motion
\be\label{effeom}
\fd{S_{eff}[x]}{x(t)}=\fd{S[x,y]}{x(t)}_{|y=y[x]}+\int_{t_t}^{t_f}dt'\fd{S[x,y[x]]}{y(t')}\fd{y[t';x]}{x(t)}=\fd{S[x,y]}{x}_{|y=y[x]}=0
\ee
shows clearly the double role the system coordinate plays in the effective dynamics: It appears twice on the list of variables of $S[x,y[x]]$, first as a virtual variational parameter to deduce forces and second as a position defining parameter and only the second role is taken up outside of the variational calculation. This is redoubling (C2) and suggests the generalization
\be
F(x,\dot x)=-\partial_xU_t(x,\dot x,x',\dot x')_{|x'=x}+\frac{d}{dt}\partial_{\dot x}U(x,\dot x,x',\dot x')_{|x'=x}
\ee
of eq. \eq{hol}.

The equation of motion obtained by eliminating dynamical degrees of freedom is always non-local in time with infinitely long non-locality. In fact, a local variation of the system coordinate at the time of the first horizontal dashed line on Fig. \ref{sectas}(a) changes the environment equation of motion. This in turn changes the environment trajectory for the rest of the time evolution, the second horizontal line of Fig. \ref{sectas}(a) can be at any later time. Hence the semi-holonomic force is defined by a non-local functional $S_{sh}[x,x']$,
\be\label{semihol}
F(t)=\fd{S_{sh}[x,x']}{x(t)}_{|x'=x}.
\ee
The structure $S[x,y[x]]$ of the effective action assures that the semi-holonomic forces cover all possible open forces existing within a subsystem of a closed dynamics.

\subsection{Action of an open system}\label{aopsyss}
To find the action and the variational trajectory space in the presence of semi-holonomic forces we write the full action in the form $S[x,y]=S_s[x]+S_e[y]+S_i[x,y]$. We shall use initial conditions for the system and seek the action corresponding to the case of either parallel or antiparallel system and environment time arrows, shown in Fig. \ref{sectas} (a) and (d), respectively.

To form an idea what we need let us rely first on perturbation expansion and the choice of a simple system-environment interaction corresponding to the term $L_i=gxy$ $g$ in the Lagrangian. In the case of a causal effective dynamics of Fig. \ref{sectas} (a) the perturbative elimination of the environment, the iterative solution of the environment equation of motion, generates the non-local retarded contribution $S^{(r)}=g^2x'(t_2)D_e^r(t_2,t_1)x'(t_1)$ to the effective action in the leading order where $t_1$ and $t_2$ correspond to the time of right and the left oriented horizontal dashed lines in Fig. \ref{sectas} (a), respectively and $D_e^r$ stands for the retarded Green function of the environment. In the case of the advanced effective dynamics, shown in Fig. \ref{sectas} (b) the leading order interaction is represented by $S^{(a)}=g^2x'(t_2)D_e^a(t_2,t_1)x'(t_1)$, $D_e^a$ being the advanced environment Green function. Thus the leading order interaction is given by the action
\be\label{intact}
S_i^{\stackrel{(r)}{(a)}}=g^2\int_{t_i}^{t_f}dt_1dt_2x'(t_2)[D_e^n(t_2,t_1)\pm D_e^f(t_2,t_1)]x'(t_1).
\ee

The need of keeping track both the retarded and advanced terms is argument (C3) of redoubling: The contribution of the far Green function is vanishing because the Green function is sandwiched between the same trajectory. This is a well known problem of the traditional STP scheme which is having difficulties in supporting interactions with odd time reversal parity. For instance the Lorentz force in electrodynamics has negative time reversal parity however it can be derived from an STP variation principle owing to its negative space inversion parity. The representation of the time inversion asymmetric part of a one dimensional harmonic force needs ``another'' system trajectory handled independently in the variation.

The action for the trajectory doublet has to be introduced in such a manner that the solution of the equation of motion brings the two trajectories to overlap. For this end we follow the procedure outlined in section \ref{atas} and introduce the action $S[\hx,\hy]=S[x_+,y_+]-S^*[x_-,y_-]$ up to the infinitesimal generalized $\epsilon$-prescription terms for the full closed system and define the variational trajectory space with the same initial conditions for the two trajectories and with the identification of the final coordinates, $x_+(t_f)=x_-(t_f)$, $y_+(t_f)=y_-(t_f)$. The derivation of the effective action follows the steps outlined above and one arrives at
\be\label{ctpeffact}
S_{eff}[\hx]=S_s[x_+]+S_e[y_+[\hx]]+S_i[x_+,y_+[\hx]]-S^*_s[x_-]-S^*_e[y_-[\hx]]-S_i[x_-,y_-[\hx]]
\ee
where
\be\label{elim}
\fd{}{\hy(t)}\{S_e[y_+]+S_i[x_+,y_+]-S^*_e[y_-]-S^*_i[x_-,y_-]\}=0.
\ee
The effective action \eq{ctpeffact} can be written as
\be\label{infleffact}
S_{eff}[\hx]=S_s[x_+]-S^*_s[x_-]+S_{infl}[x_+,x_-],
\ee
the sum of the closed system CTP action and the influence functional \cite{feynman}
\be\label{inflact}
S_{infl}[\hx]=S_e[y_+[\hx]]+S_i[x_+,y_+[\hx]]-S^*_e[y_-[\hx]]-S^*_i[x_-,y_-[\hx]]
\ee
representing the effective interactions. The simplest open system, a damped harmonic oscillator, corresponds to the Lagrangian
\be\label{dho}
L=\frac{m}2[\dot x_+^2-\dot x_-^2-\omega_0^2(x_+^2-x_-^2)+\nu(\dot x_+x_--x_+\dot x_-)],
\ee
with the equation of motion \cite{bateman}
\be
\ddot x_\pm=-\omega_0^2x_\pm-\nu\dot x_\mp.
\ee
The quantum Brownian motion, a particle interacting with infinitely many harmonic oscillators can be treated in a similar manner \cite{caldeira}. The model is simple enough to be solved both within the STP and the CTP schemes using equations of motion and effective action, respectively.

Several remarks are in order at this point:

(1) One can now better understand the necessity of the time inversion performed at $t_f$ in the intuitively introduced scheme of section \ref{atas}: It yields the opposite sign in front of the action of the two CTP doublet trajectories, needed to retain the contribution of the far Green function term in eq. \eq{intact}.

(2) There is a clear difference between closed and open dynamics: While the action \eq{cctpact} of a closed dynamics has no finite $\ord\epsilon$ coupling between the copies $x_+$ and $x_-$ the closing of the final environment coordinates $y_+(t_f)=y_-(t_f)$ during the elimination \eq{elim} couples $x_+$ and $x_-$ with a finite strength.

\begin{figure}
\includegraphics[scale=.4]{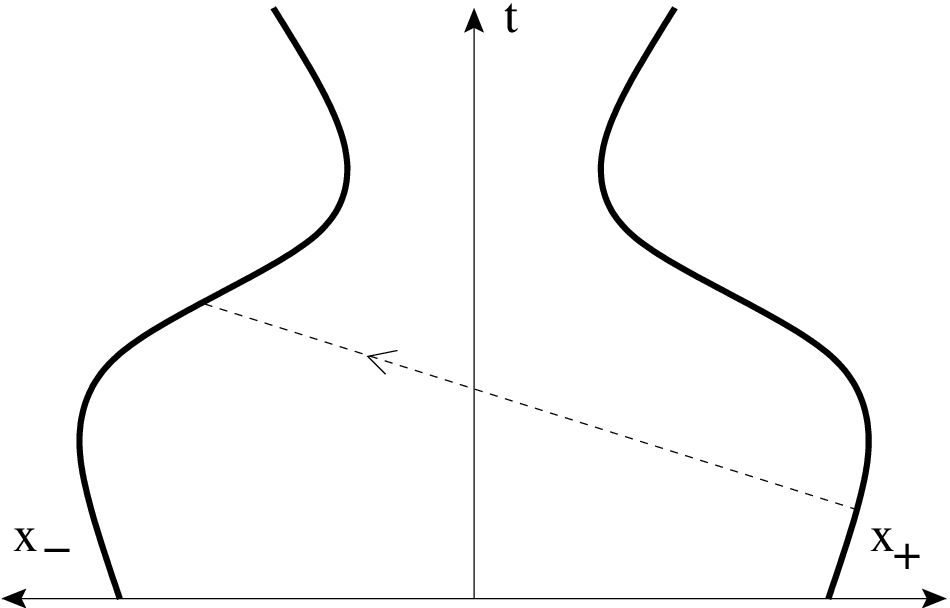}\hskip.5cm
\includegraphics[scale=.4]{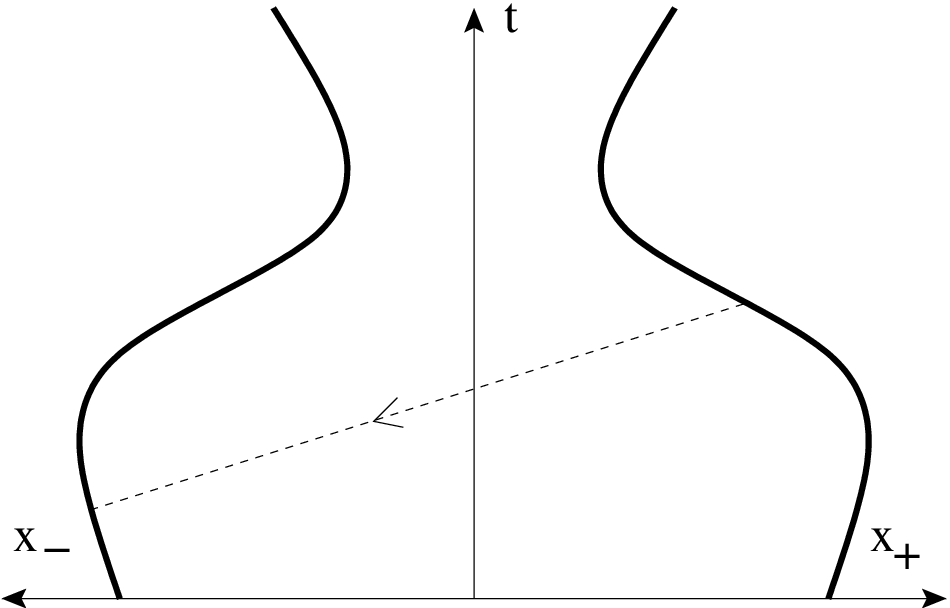}

(a)\hskip7cm(b)
\caption{An open interaction channel of $x_+$ with the environment environment is represented as an interaction with $x_-$. The dashed line stands for the environment induced self interaction and (a): $\tau_e=1$; (b): $\tau_e=-1$.}\label{dfpath}
\end{figure}

(3) The effective action can be written by separating the STP and the genuine CTP terms,
\be\label{clopch}
S_{eff}[\hx]=S_1[x_+]-S^*_1[x_-]+S_2[\hx]
\ee
where $S_1$ stands for the contributions of exclusively either $x_+$ or $x_-$ and $S_2$ contains the rest, the terms coupling $x_+$ and $x_-$. The difference between the splitting \eq{inflact} and  \eq{clopch} of the effective action is that while the influence functional contains all, closed and open, effective interactions, eg. mass renormalization and the Abraham-Lorentz force in the effective the dynamics of a single electron, $S_2$ includes only the open interaction channels. The role of $S_1$ and $S_2$ can be revealed by the help of the parametrization $x_\pm=x\pm x_d/2$: The equation of motion arising from the variation of $x$ by ignoring the infinitesimal imaginary terms,
\be\label{triveom}
\fd{S_1[x+\frac{x_d}2]}{x}-\fd{S^*_1[x-\frac{x_d}2]}{x}+\fd{S_2[x+\frac{x_d}2,x-\frac{x_d}2]}{x}=0,
\ee
is trivially satisfied since $x_+(t)-x_-(t)=x_d(t)=0$ after imposing the equations of motion by construction. The variation of $x_d$ yields the physical equation of motion,
\be
\fd{S_1[x]}{x}+\fd{S_2[x_+,x_-]}{x_+}_{|x_+=x_-=x}=0,
\ee
and shows that the holonomic and the semi-holonomic forces arise from $S_1$ and $S_2$, respectively, shown in Fig. \ref{dfpath} where the dashed line stands to the environment far Green function. The interactions are separated into two classes by the expressions \eq{infleffact} and \eq{clopch} the difference being that the system interaction is defined by the original system dynamics ($S_s$) in the former and by all closed system interactions ($S_1$) in the latter. The classification of \eq{clopch} is more natural since the separation of the full system into the observed subsystem and its environment is introduced only by us.

(4) The time reversal can be realized in two different manners in open systems: The full time reversal acts on both the system and its environment. This is a formal, trivial reparametrization invariance, expressed in the form of eq. \eq{ctpsym}. The partial time reversal effects is performed only on the observed system and is realized as a full time reversal followed by the transformation $t\to-t$ in the system trajectories and a complex conjugation of the corresponding action terms, $S_s[x_\pm]\to -S^*_s[x_\mp]$, $S_i[x_\pm,y]\to -S^*_i[x_\mp,y]$. This definition confirms that the interaction with the environment breaks the time reversal invariance of the effective system dynamics.

The final two remarks concern difficulties posed by the non-local nature of the action.

(5) The necessary auxiliary conditions for a generic non-local equation of motion are usually unknown. Nevertheless the construction of the effective action assures that both $\delta S_{eff}[\hx]/\delta\hx(t)=0$ and $\delta S[\hx,\hy]/\delta\hx(t)=0$ need the same initial conditions for given environment auxiliary conditions, a natural expectation from the open dynamics. Hence the original system phase space remains the manifold of possible auxiliary conditions of the open dynamics. This is necessary to make the elimination procedure of the environment degrees of freedom as described after eq. \eq{hol} possible.

(6) It is difficult to calculate $S_{eff}[\hx]$ and to find the solutions of a non-local equation of motion hence we have to rely on some approximation. Different restrictions apply to an acceptable approximation. (i) While the equation of motion derived from the exact $S_{eff}[\hx]$ needs just the physical auxiliary conditions this does not usually hold when an approximation is employed: The auxiliary conditions of a generic integro-differential equations are usually unknown. The usual approximation scheme, the Landu-Ginzburg double expansion in the derivative and the coordinate, leads to equations of motion with higher order time derivatives which in turn require further unphysical auxiliary conditions. (ii) Noether's theorem establishes the usual relation between symmetries and conservation laws for full closed dynamics of $S[\hx,\hy]$ but the conservation laws related to symmetries which are violated when performed only on the system are lost in the open dynamics. The recovery of the proper conservation laws of a non-local dynamics is made non-trivial owing to non-local conserved quantities even without symmetry \cite{nonloc}. (iii) The stability of the dynamics of $S[\hx,\hy]$ is not sufficient to assure the stability of the approximate open equation of motion: The higher order time derivatives in a local equation of motion obtained in the derivative expansion make the dynamics unstable \cite{ostrogadsky}. The stability of a generic non-local dynamics is out of control owing to the unboundendedness from below of the conserved quantity corresponding to the time translation invariance \cite{nonloc}.

\section{Ancilla}\label{ancillas}
The argument (C3) suggests that the CTP copy of the system actually represents the environment. We elaborate now on this point of view and seek a non-trivial extension of the original variational principle to cover open systems by introducing the copy as a simple, efficient ancilla.

\subsection{Environment as a copy}
If a small system is interacting with a large environment then it is not necessary to possess the full information about the environment to find its imprint on the dynamics of the system. The selection and the representation of the necessary information can be achieved by introducing a sufficiently simple ancilla as a new, reduced environment. How to choose the ancilla and its interaction with the system?

This question can be posed in a more illuminating manner by looking into the conservation laws whose violation is the hallmark of open dynamics. The energy is decreased by a Newtonian friction force which is proportional to the velocity despite the explicit time translation invariance of the equation of motion. How can we modify the variational principle in order to reproduce such a non-conservative force and the resulting energy loss without breaking time translation symmetry of the effective dynamics?

The answer to this question is obvious, the ancilla should absorb the lost energy. Rather than trying to circumvent the second law of thermodynamics and to accumulate the dissipated energy in a small, ordered ancilla we can arrive at a solution in two simple steps. First, it is trivial to compare the energy stored in the observed system and the ancilla if the latter is chosen to be a copy of the former. This is redoubling (C4).

As a side remark, such an {\em ancilla = observed system} construction is optimal for the representation of the effective dynamics since the environment is reduced to an ancilla with the same complexity as the system. The second step is to realize that the energy conservation can trivially be achieved by assuring that the energy is defined with the opposite sign in the original system and its copy. The ancilla is an ``anti-system''. Since the conserved quantities are linear in the Lagrangian according to the Noether theorem the Lagrangian of the full system is chosen to be antisymmetric with respect to the exchange of the system and the ancilla as in eq. \eq{inflact}.

\subsection{Noether theorem and semi-holonomic forces}
The procedure can easily be demonstrated by the derivation of the Noether theorem for the Lagrangian $L=L_1(x_+,\dot x_+,t)-L_1(x_-,\dot x_-,t)+L_2(x_+,\dot x_+,x_-,\dot x_-,t)$. The momentum balance equation arises by performing an infinitesimal translation in the coordinate space, $\hx\to\hx+\hat\xi$ and by treating the time dependent $\delta\hx=\hat\xi$ with $\hat\xi(t_i)=\hat\xi(t_f)=0$ as a special variation. We have now the two dimensional parameter $\xi_\sigma$ to describe a variation with the linearized action
\be\label{xiaction}
S[\hat\xi]=\int dt\hat\xi\left(\frac{d}{dt}\pd{L}{\dot{\hx}}-\pd{L}{\hx}\right)
\ee
and there are two independent variations: the system and its ancilla can be varied in the same or in the opposite ways. The linearized action is independent of $\xi=(\xi_++\xi_-)/2$ according to \eq{triveom} indicating that ``total momentum'', defined by the identical variation, is vanishing up to $i\ord\epsilon$ terms in agreement with the remark that the momentum of the two copies is defined with the opposite sign. The Noether theorem corresponding to the opposite variation of the two trajectories, the equation of motion for $\xi_d=\xi_+-\xi_-$, adds rather than subtracts the contributions of the two momenta and is actually a balance equation,
\be
\frac{d}{dt}\pd{L_1}{\dot x}=\pd{L_1}{x}+\left(\pd{L_2}{x_+}-\frac{d}{dt}\pd{L_2}{\dot x_+}\right)_{|\dot x_\pm=\dot x,x_\pm=x}.
\ee
The change of the momentum defined by the help of the STP part of the Lagrangian is due to the violation of the translation invariance of the closed dynamics and the semi-holonomic forces.

It is instructive to introduce the renormalized momentum
\be
p_r=\pd{L_1}{\dot x}+\pd{L_2}{\dot x_+}_{|\dot x_\pm=\dot x,x_\pm=x}
\ee
where the second term represents the image of the system, the ``polarization''  of the environment which is generated by the system-environment interactions and moves together with the system. Its balance equation,
\be\label{momcons}
\dot p_r=\pd{L_1}{x}+\pd{L_2}{x_+}_{|\dot x_\pm=\dot x,x_\pm=x}.
\ee
shows that the renormalized momentum is non-conserved owing to the breakdown of the translation invariance of the system alone without shifting its environment. By assuming a translation invariant full closed dynamics for the system and its environment the second term is non-vanishing if the interaction or the environment initial conditions are non-symmetric.

The balance equation for angular momentum comes from a linear transformation of the multi-component coordinate vector, $\delta x_\sigma=\xi_\sigma\tau x_\sigma$ where $\tau$ stands for a generator of the rotation group. The linearized action of the infinitesimal angle $\xi$,
\be
S[\hat\xi]=-\sum_\sigma\int dt\left[\xi_\sigma\pd{L}{x_\sigma}\tau x_\sigma+\pd{L}{\dot x_\sigma}(\xi_\sigma\tau x_\sigma+\xi_\sigma\tau\dot x_\sigma)\right]
\ee
results the equation of motion for $\xi_d=\xi_+-\xi_-$
\be\label{angmombe}
\frac{d}{dt}L_r=x\tau\pd{L_1}{x}+\dot x\tau\pd{L_1}{\dot x}+x\tau\left(\pd{L_2}{x_+}+\dot x\tau\pd{L_2}{\dot x_+}\right)_{|\dot x_\pm=\dot x,x_\pm=x}
\ee
for the renormalized angular momentum $L_r=x\tau p_r$.

The energy equation is obtained by performing an infinitesimal translation in time, $x_\pm(t)\to x_\pm(t-\xi_\pm)$, $\xi_\pm(t_i)=\xi_\pm(t_f)=0$, and treating $\delta x_\sigma=-\xi_\sigma\dot x_\sigma$ as a variation with the liearized action
\be
S[\hat\xi]=-\sum_\sigma\int dt\left(\xi_\sigma\dot x_\sigma\pd{L}{x_\sigma}+\frac{d}{dt}(\xi_\sigma\dot x_\sigma)\pd{L}{\dot x_\sigma}\right).
\ee
The triviality of the equation of motion for $\xi$ renders the linearized action $\xi$-independent and its $\xi_d$-dependence,
\bea
S[\xi_d]&=&\int dt\left[\xi_d\left(\partial_tL_1-\frac{d}{dt}L_1\right)-\dot\xi_d\dot x\pd{L_1}{\dot x}\right]\nn
&&-\hf\sum_\sigma\sigma\int dt\left(\xi_d\dot x_\sigma\pd{L_2}{x_\sigma}+\dot\xi_d\dot x_\sigma\pd{L_2}{\dot x_\sigma}+\xi_d\ddot x_\sigma\pd{L_2}{\dot x_\sigma}\right),
\eea
is obtained by the help of the relation
\be\label{legy}
\frac{d}{dt}L_1=\dot x\pd{L_1}{x}+\ddot x\pd{L_1}{\dot x}+\partial_tL_1.
\ee
The equation of motion,
\be
\frac{d}{dt}\left(\dot x\pd{L_1}{\dot x}-L_1\right)=-\partial_tL_1+\dot x\left(\pd{L_2}{x_+}-\frac{d}{dt}\pd{L_2}{\dot x_+}\right)_{|\dot x_\pm=\dot x,x_\pm=x}
\ee
the energy balance equation, shows that the energy defined by the closed dynamics is changed by the explicit time-dependence of the closed dynamics and the work of the semi-holonomic forces.

\subsection{Generalized ancilla}
The equations of motion of closed systems are restricted by the variational principle, they are the canonical equations of classical mechanics. Open equations of motion are non-conservative hence non-canonical. We managed to extend the variation principle over open system. Does the resulting wider set of variational equations of motion define an extension of the canonical structure? It is easy to see that such an extension is without structure, more precisely an ancilla can be introduced in such a manner that any equation of motion can be derived from the extended variational principle.

Let us assume that the equation of motion for the coordinate $x$ at the time $t$ can be written in the form $F_t[x]=0$ where $F_t[x]$ is an arbitrary functional of the trajectory $x(t)$. We introduce another coordinate $x_d$ and define the action
\be
S_F[x,x_d]=\int dt'x_d(t')F_{t'}[x]+S'[x_d]
\ee
for the two trajectories where $S'$ is an arbitrary odd functional, $S'[x_d]=-S'[x_d]$. It easy to see that the action
\be
S[\hx]=S_F\left[\hf(x_++x_-),x_+-x_-\right]
\ee
equipped with the generalized $\epsilon$-prescription terms generates the variational equation $F_t[x]=0$ and $x_d=0$ within the variational space of section \ref{atas}. Argument (C5) for redoubling is that it leads to an ancilla in such a manner that the equation of motion can be derived by the variational principle.

\section{Closed quantum dynamics}\label{clqds}
The first indication of the redoubling in quantum mechanics, argument (Q1), is the way the time arrow is introduced. The auxiliary conditions for the time reversal invariant Newton equation, a second order differential equation, break the time reversal invariance and can be used to orient the time for the classical motions. The Hamilton equation is first order but its auxiliary condition remains non-invariant under time reversal since the canonical pairs have opposite time reversal parities. The time evolution of a quantum state, a continuous trajectory in a linear space, can be defined by its generators hence the equation of motion is first order in quantum mechanics. How to encode the direction of time within a state?

The solution of this problem is well known, it is the introduction of an internal time reversal parity for the quantum states. The usual procedure is to use positive time reversal parity coordinate eigenstates and to define the wave function $\psi(x)=\la x|\psi\ra$ of the time reversed state by complex conjugation,  $T\psi(x)=\la x|T|\psi\ra=\la x|\psi\ra^*=\la\psi|x\ra$. Hence the representation of the direction of time requires the simultaneous use of bras and kets in the expectation value of observables.

The structure of a probability assignments in Hilbert spaces provides argument (Q2) for redoubling. In fact, the physical state, a well defined probability assignment, can be represented by a density matrix $\rho$, an element of the Liouville space of operators acting on the Hilbert space of pure states. The pure states, $\rho=|\psi\ra\la\psi|$, are actually composed by the equivalent bra and ket.

The expectation value of the coordinate dependent operator $A(x)$ at time $t$ is given by
\bea\label{pintexpval}
\la\psi_i|U^\dagger(t,t_i)AU(t,t_i)|\psi_i\ra&=&\int d\hx_idx_fA(x_f)\la x_{i+}|\psi_i\ra\la\psi_i|x_{i-}\ra\nn
&&\times\int_{\hx(t_i)=\hx_i}^{x_\pm(t)=x_f}D[\hx]e^{\ih(S[x_+]-S^*[x_-])}
\eea
in the path integral representation. Therefore the integration over the trajectories $x_+(t)$ and $x_-(t)$ of Fig. \ref{ctppathf} represents the quantum fluctuations within $U$ and $U^\dagger$ and the action is given by eq. \eq{cctpactf}.

The expression \eq{pintexpval} of the reduced density matrix is a natural starting point to understand the historical (Q0) argument for the CTP formalism. The goal was to work out the perturbation expansion in the Heisenberg representation for observables \cite{schw}, for the causal Green functions \cite{keldysh} and at finite temperature \cite{umezawa}. The redoubling arises from the independent perturbation series of the time evolution operators $U(t)$ and $U^\dagger(t)$.

\section{Open quantum dynamics}\label{oquantms}
The quantum fluctuations are defined in a basis dependent manner and are described by a complex valued function $\psi_n=\la n|\psi\ra$ for a basis set $\{|n\ra\}$ in a pure state $|\psi\ra$. To cover the mixed states the quantum fluctuations are generalized for a density matrix $\rho$ as $\rho_{n_+,n_-}=\la n_+|\rho|n_-\ra$. It is a remarkable fact while that the quantum fluctuations in the bra and the ket components are independent in a pure state $|\psi\ra$, $\rho_{n_+,n_-}=\psi_{n_+}\psi^*_{n_-}$ they become correlated in a mixed state
\be\label{mixedstate}
\rho=\sum_n|\psi_n\ra p_n\la\psi_n|
\ee
where the sum extends over more than a single linearly independent states.

In the case of a closed classical dynamics it is sufficient to solve the CTP Euler-Lagrange equation either for $x_+$ or $x_-$ to find the time evolution. When the dynamics is open and the two trajectories are coupled and one has to solve the equation of motion for both of them. The situation is similar in quantum dynamics where the ket and the bra of a pure state evolves independently in a closed dynamics and it is enough to solve the Schrödinger equation for one of them because the bra and the ket are related by time inversion. Argument (Q3) for redoubling comes from the necessity of the simultaneous handling of the quantum fluctuations of the bra and the ket factors in the density matrix, c.f. eq. \eq{otprdm} below. The coupling between the two doublers represent the open interaction channels in both the classical and the quantum cases.

The open quantum dynamics is defined in the path integral formalism by the effective bare action, introduced below, followed by a brief discussion of two characteristic features of open quantum dynamics, the emergence of a finite life-time and the semiclassical limit.

\subsection{Quantum effective actions}\label{quanteffacts}
The construction of the effective quantum action is based on the reduced density matrix of the coordinate $x$ within a closed system of section \ref{aopsyss}
\bea\label{reddensm}
\rho(t,\hx)&=&\la x_+|\Tr_e[U(t,t_i)\rho_iU^\dagger(t,t_i)]|x_-\ra\nn
&=&\int d\hx_id\hy_idy_f\rho_i(x_{i+},y_{i+},x_{i-},y_{i-})\nn
&&\times\int_{\hx(t_i)=\hx_i,\hy(t_i)=\hy_i}^{\hx(t)=\hx,y_\pm(t)=y_f}D[\hx]D[\hy]e^{\ih(S[x_+,y_+]-S^*[x_-,y_-])},
\eea
where the integration of the final coordinate of the environment trajectory stands for the trace $\Tr_e$ over the environment Hilbert space. One introduces the influence functional by the equation
\bea\label{qinflf}
e^{\ih S_{infl}[\hx]}&=&\int d\hy_idy_f\rho_e(y_{i+},y_{i-})\nn
&&\int_{\hy(t_i)=\hy_i}^{y_\pm(t)=y_f}D[\hy]e^{\ih(S_e[y_+]+S_i[x_+,y_+]-S_e^*[y_-]-S_i^*[x_-,y_-])},
\eea
where the initial density matrix is assumed to be factorizable, $\rho_i(x_{i+},y_{i+},x_{i-},y_{i-})=\rho_s(x_{i+},x_{i-})\rho_e(y_{i+},y_{i-})$ and rewrite the reduced density matrix as
\be\label{otprdm}
\rho(t,\hx)=\int d\hx_id\hy_idy_f\rho_s(x_{i+},x_{i-})\int_{\hx(t_i)=\hx_i}^{\hx(t)=\hx}D[\hx]e^{\ih S_{eff}[\hx]},
\ee
where the bare effective action $S_{eff}$ is given by the right hand side of \eq{infleffact}. The system and the environment are treated in the CTP scheme with identical final points of the doublet trajectories and in the Open Time Path (OTP) scheme with different, fixed end points, respectively.

The relation with the classical dynamics can be found by the help of the generator functional
\bea\label{genfunct}
e^{\ih W[\hj]}&=&\Tr[U(t,t_i;j_+)\rho_iU^\dagger(t,t_i;-j_i)]\nn
&=&\int D[\hx]e^{\ih S[\hx]+\ih\int dt\hj(t)\hx(t)}
\eea
for the connected Green functions defined by using \eq{cgrfnct} for $W[\hj]$ rather than $Z[\hj]$ where $U(t,t_i;j)$ is the time evolution operator for the closed dynamics of the system and its environment in the presence of the external source $j$ coupled to $x$. The inverse functional Legendre transform $W[\hj]\to S[\hx]$ obtained by eqs. \eq{legendre} and \eq{ilegendrev} defines the quantum effective action $\Gamma[\hx]$. The trajectory $\hx(t)$ defined by \eq{ilegendrev} at $\hj=0$ yields the physical trajectory
\be\label{xexpval}
x(t)=\int D[\hx]e^{\ih S_{eff}[\hx]}x_\pm(t)=\la x(t)\ra
\ee
which satisfies the variational equation of $\Gamma[\hx]$ according to eq. \eq{legendrev}. Hence the quantum effective action defined is the analogy of the classical action since the former yields equation of motion for the expectation value $\la x\ra=\la x_\pm\ra$ and the latter for the classical coordinate $x$.

It is instructive to consider the case of an open harmonic oscillator where the second order Green functions are identical in the classical and the quantum case owing to Ehrenfest's theorem. The natural proposal for the action is \eq{genepspr} finite $\epsilon$ however it produces an $\ord\epsilon$ acausal contribution in $D^r$. Instead one seeks the most general quadratic Lagrangian compatible with the full time inversion symmetry \eq{ctpsym} \cite{ines}
\bea\label{hod}
L&=&\frac{m}2(\dot x_+^2-\dot x_-^2)-\frac{m\omega^2}2(x_+^2-x_-^2)+\frac{m\nu}2(\dot x_+x_--\dot x_-x_+)\nn
&&+\frac{i}2[d_0(x_+-x_-)^2+d_2(\dot x_+-\dot x_-)^2].
\eea
producing $K^n=m(\omega^2-\omega_0^2)$, $K^i=d_0+d_2\omega^2$, and $K^f=im\nu\omega$ which give
\bea
D^{\stackrel{r}{a}}&=&\frac1{m[\omega^2-\omega_0^2\pm i\omega\nu]},\nn
D^i&=&-\frac{d_0+d_2\omega^2}{m[(\omega^2-\omega_0^2)^2+\omega^2\nu^2]},
\eea
where the Heaviside function is smeared. The closed limit is $\nu=\epsilon/\omega_0$, $d_0=m\epsilon$, and $d_2=0$. This Lagrangian is better suited for phenomenological applications because $\nu$, $d_0$ and $d_2$ may be finite without acausality. The open oscillator experiences a Newton friction force $F_f=-m\nu\dot x$ and the parameters $d_0$ and $d_2$ which drop out from the classical equation of motion control decoherence in the coordinate basis \cite{dyndec}.

\subsection{Finite life-time and preservation of the total probability}
The open interaction channels lead to a more radical changes in the quantum dynamics than in the classical case, the parameters of the action may acquire imaginary part. In fact, the parameters of the bare and the effective quantum action are given in terms of connected and one-particle irreducible Greens functions, respectively, evaluated at appropriately chosen kinematical point. The Green functions may assume complex values when an internal line of a contributing Feynman graph is on the mass-shell in agreement with the optical theorem. Hence these new parameters of the theory, the imaginary parts, parametrize diffusive processes such as finite line widths.

The system-environment separation is usually realized by starting with a closed dynamics within a direct product Hilbert space ${\cal H}_t={\cal H}_s\otimes{\cal H}_e$ where the factors ${\cal H}_s$ and ${\cal H}_e$ stand for the linear space of the system and the environment, respectively. One should bear in mind that this structure is not necessary. In fact, the definition of the effective action \eq{infleffact} is obtained in eq. \eq{otprdm} by integrating out the environment modes whose definition is given by some separation of the integral variables of the closed dynamics into two classes without any relation to the Hilbert space structure. This possibility explains the open effective dynamics of composite operators, such as the electric density and current in QED \cite{saso}.

While the open dynamics is non-unitary it preserves the total probability.  In fact, the system and its environment together obey a closed unitary dynamics hence $W[\hj]=0$ for physical sources, $j_+=-j_-$, implying $\Tr\rho_s=1$ for the reduced system density matrix, c.f. eq. \eq{genfunct}. When a state has a finite life-time it leaks into the environment and induces mixed states. In other words, a state with a  finite life-time can not stay pure.

\subsection{Semiclassical limit}
A natural application of a CQCO formalism is the classical limit. It is a widespread view that in phenomenons where $\hbar$ can be treated as a small parameter the quantum effects are weak when some macroscopic quantum effects are ignored. the usual argument goes by referring to the $\ord\hbar$ Heisenberg commutation relation or by pointing out that the traditional path integral expression for the transition amplitude between coordinate eigenstates
\be\label{tradpint}
\la x_f|e^{-\ih Ht}|x_i\ra=\int_{x(0)=x_i}^{x(t)=x_f}D[x]e^{\ih S[x]}
\ee
is dominated by the classical trajectory as $\hbar\to0$. Such a view is too naive \cite{ballantine,klein} and the CQCO scheme offers a more realistic treatment of the classical limit.

Though the sufficient conditions for the classical limit remain unknown there are few well known necessary conditions, such as the decoherence, the suppression of the interference between macroscopically different states, and the return of the determinism, the narrowing of the probability distributions for the observables. The first condition excludes closed systems where the time evolution is unitary and their fully resolved dynamics remains forever quantum. The second condition can be satisfied if the observable is the macroscopic average of microscopic quantities according to the central limit theorem \cite{macr}.

The conflict between the simplistic $\hbar\to0$ condition and the more involved arguments about the need of the decoherence and the macroscopic limit can partially be understood by comparing the path integrals \eq{otprdm} and \eq{tradpint}. On the one hand, the decoherence in the coordinate basis consists of the suppression of trajectory pairs with $x_d=x_+-x_-$ on the macroscopic scale in \eq{otprdm}. Hence the trajectory pairs with $x_d\sim0$ dominate the path integral and the two copies ''stick together`` in the classical limit yielding dominantly imaginary action $S_{eff}[x,x]$, c.f. eq. \eq{ctpsym}, for the common trajectory $x=x_+=x_-$. The imaginary part of the effective action arises from the system-environment interactions hence it is  assumed to be small and the dependence of the integrand of the functional integral \eq{otprdm} on the trajectories $x_+\approx x_-$ is weak for weak system-environment interactions. The fluctuations in the path integral with approximately constant integrand are large and the dynamics of $x(t)$ is soft. On the other hand, the formal limit $\hbar\to0$ localizes the dominant contributions in \eq{tradpint} around the classical trajectory hence the fluctuations are small and the dynamics is hard. Is the semiclassical dynamics soft or hard?

The answer depends on the auxiliary conditions. In a realistic case the experimentalists control the initial conditions only. The soft path integral \eq{otprdm} reflects the easy excitability of the environment owing to its dense excitation spectrum. If a pure initial and final states of an isolated macroscopic system are fixed in a thought experiment then the action is large compared to $\hbar$ and the transition amplitude \eq{tradpint} is dominated by the classical saddle point of a hard dynamics.

But notice that \eq{tradpint}, a transition amplitude between pure states in a closed dynamics of a macroscopic system, can not be measured in practise. The formal argument about the dominance of the classical trajectory may remains valid only for very short time in agreement with the general expectation that short time, high energy motion is semiclassical. However the decoherence which builds up in during a long time evolution invalidates this argument.

\section{Summary}\label{summs}
To understand realistic mechanical systems from first principles we need a CQCO formalism which is equally applicable for classical, quantum, closed and open dynamics. This condition is satisfied by the CTP scheme where the dynamics can be defined by an action functional. However such a wide applicability is based on an unusual feature, a formal redoubling of degrees of freedom. Some possible origins of redoubling were presented in the present work. Though these arguments are obviously related mathematically they correspond to different physical issues.

The arguments (C1) and (Q1) originate from the existence of a time arrow in closed dynamics. The structure of classical semi-holonomic forces and of the probability assignments in Hilbert spaces leads to arguments (C2) and (Q2), respectively. The action governing the effective open dynamics gives rise to arguments (C3) and (Q3). The arguments (C4) and (C5) are found by viewing the environment as an ancilla which absorbs the violation of the conservation laws arising from the Noether theorem and allows to transform an arbitrary effective equation of motion into a canonical form. Finally, the historical motivation of redoubling, argument (Q0), is to provide a perturbation expansion for obervables in Heisenberg representation and for the causal Green functions.

The attention was restricted to quantum mechanics in this work. The generalization of these ideas for field theory is straightforward and the importance of the redoubling can bee seen from the following two remarks. The open interaction channels in gauge theories and General Relativity introduce completely new gauge invariant structures, such as Wilson loops or lines and topological quantities arising from connecting two field configurations. Thus the classical action receives qualitatively new terms in case of unobserved charges. Quantum field theories represent closed dynamics only if the UV cutoff is kept fixed as a physical parameter. But as soon as we try to hide it by an appropriate redefinition of the bare parameters we open the dynamics, the UV modes become the environment. Hence any quantum field theory without explicit UV cutoff must be presented with redoubled field which introduces new parameters in the action \cite{openqft}.
 
The redoubling makes a wide class of physical phenomenons accessible and offers new points of view suggesting that it should be included into our standard tool box of theoretical physics.

\appendix
\section{Two-point Green function}\label{hogfa}
The most important Green function with two legs, $\hD(x_1,t_2)$, is derived and its structure commented in this appendix.

\subsection{Classical closed harmonic oscillator}
The Green function of the CTP formalism is calculated below for a classical harmonic oscillator, defined by the Lagrangian $L=m(\dot x^2-\omega_0^2x^2)/2$. Since the external source $j$ is supposed to generate the non-trivial dynamics the trivial initial conditions with vanishing coordinate an velocity are assumed. To facilitate the inversion of the kernel of the action we discretize the time interval $-T<t<0$ by introducing $t_n=n\dt-T$, $n=0,\ldots,N-1$, $\dt=T/N$ and write the CTP action in the form
\be
S=\frac{m}2\sum_{\sigma=\pm}\sigma\sum_{n=0}^{N-1}\left[\frac{(x_{n+1,\sigma}-x_{n,\sigma})^2}\dt-\dt\omega_0^2x_{n,\sigma}^2\right].
\ee
The rule of the partial integration with the finite difference operators $\nabla^+f_n=f_{n+1}-f_n$, $\nabla^-f_n=f_n-f_{n-1}$,
\be
\sum_{n=1}^{N-1}\nabla^+f_ng_n=-\sum_{n=1}^{N-1}f_n\nabla^-g_n+f_Ng_{N-1}-f_1g_0,
\ee
yields the action
\be
S=\hf\sum_{\sigma,\sigma'}\sum_{n,n'=1}^{N-1}x_{n,\sigma}D^{-1}_{0(n,\sigma),(n'\sigma')}x_{n',\sigma'}+\sum_\sigma\sum_{n=1}^{N-1}x_{n,\sigma}B_{n,\sigma}z
\ee
for two separate trajectory segments $(x_{0,\pm},\ldots,x_{N-1,\pm})$ which become the CTP trajectory doublet with the common final point $z=x_{N,\pm}$. The Green function $\hD_0$ is defined by
\be
D^{-1}_{0(n,\sigma),(n'\sigma')}=-\delta_{\sigma,\sigma'}m\left[\sigma\left(\frac1\dt\Delta_{n,n'}+\dt\omega_0^2\delta_{n,n'}\right)-i\dt\epsilon\delta_{n,n'}\right]
\ee
with $\Delta_{n,n'}=(\nabla^-\nabla^+)_{n,n'}=\delta_{n,n'+1}+\delta_{n,n'-1}-2\delta_{n,n'}$ and describes the response to the external source within the time interval $-T<t<0$ with Dirichlet boundary condition. The time-dependent internal source
\be
B^\sigma_t=-\delta_{t,T-\dt}\frac{\sigma m}{\dt}
\ee
represents the common end point. The calculation of $\hD_0$ is easiest by the help of the complete set of eigenfunctions of $\hD^{-1}_0$,
\be
\phi_n(t)=\sqrt{\frac2{T}}\sin\pi\frac{t}{T}n,
\ee
resulting
\be
D_{0++}(t,t')=\frac{2}{Tm}\sum_{n=1}^N\frac{\sin\omega_nt\sin\omega_nt'}{\hat\omega^2_n-\omega_0^2+i\epsilon}
\ee
where $\hat\omega_n=\frac2\dt\sin\frac{\pi}{2N}n$ and $\omega_n=\frac\pi{T}n$. The continuum limit $\dt\to0$ is straightforward,
\bea\label{dpp0}
D_{0++}(t,t')&=&-\frac1{2Tm}\sum_{n=1}^N\frac{(e^{i\frac\pi{T}nt}-e^{-i\frac\pi{T}nt})(e^{i\frac\pi{T}nt'}-e^{-i\frac\pi{T}nt'})}{\frac4{\dt^2}\sin^2\pi\frac{\dt n}{2T}-\omega_0^2+i\epsilon}\nn
&=&\frac1{4\pi m}\int_{-\infty}^\infty d\omega\frac{e^{i\omega(t-t')}-e^{i\omega(t+t')}}
{(\omega-\omega_0+i\epsilon)(\omega+\omega_0-i\epsilon)}\nn
&=&\frac{i}{2m\omega_0}[e^{i(\omega_0-i\epsilon)(t+t')}-e^{-i(\omega_0-i\epsilon)|t-t'|}].
\eea

The full CTP Green function can be obtained by performing the Legendre transformation from the action $S=\hx\hD^{-1}\hx/2+\hj\hx$ written in condensed vector notation by suppressing the indices to the generator functional
\be
W=\hf\hx\hD_0^{-1}\hx+\hx(\hat Bz+\hj)
\ee
where the source $j$ is coupled to the CTP trajectories. The coordinates are eliminated in two steps, first the CTP trajectories are replaced by the solution $\hx=-\hD_0(\hat Bz+\hj)$ leading to
\be
W=-\hf(z\hat B+\hj)\hD_0(\hat Bz+\hj).
\ee
This is followed by the elimination of the common end point, $z=-\hat B\hD_0\hj/\hat B\hD_0\hat B$, resulting $W=-\hj\hD\hj/2$ with
\be\label{ctpgfnt}
\hD=\hD_0-\hD_0\hat B\frac1{\hat B\hD_0\hat B}\hat B\hD_0.
\ee

The last term on the right hand side describes the impact of the coupling of the separate trajectory segments by the common end point and contains
\bea
D_{0++}(t,-\dt)&=&-\frac{2}{Tm}\sum_{n=1}^N\frac{\sin\frac\pi{T}nt\sin\frac\pi{T}n\dt}{(\frac\pi{T}n)^2-\omega_0^2+i\epsilon}\nn
&=&\frac{i}{Tm}\sum_{n=1}^N\frac{(e^{i\frac\pi{T}nt}-e^{-i\frac\pi{T}nt})\frac\pi{T}n\dt}{(\frac\pi{T}n)^2-\omega_0^2+i\epsilon}\nn
&=&\frac{i\dt}{\pi m}\int_{-\infty}^\infty d\omega\frac{\omega e^{i\omega t}}{(\omega-\omega_0+i\epsilon)(\omega+\omega_0-i\epsilon)}\nn
&=&-\frac{\dt}{m}e^{i(\omega_0-i\epsilon)t}
\eea
and
\bea
D_{0++}(-\dt,-\dt)&=&\frac2{Tm}\sum_{n=1}^N\frac{\sin^2\frac\pi{T}n\dt}{\frac{4}{\dt^2}\sin^2\pi\frac{\dt n}{2T}-\omega_0^2+i\epsilon}\nn
&=&\frac{2\dt^2}{mT}\sum_{n=1}^N\frac{\frac4{\dt^2}\sin^2\frac{\dt\omega_n}2(1-\sin^2\frac{\dt\omega_n}2)}{\frac4{\dt^2}\sin^2\frac{\dt\omega_n}2-\omega_0^2+i\epsilon}.
\eea
It is advantageous to split the latter sum into two parts,
\bea
D_{0++}(-\dt,-\dt)&=&\frac{2\dt^2}{mT}\sum_{n=1}^N\left(1-\sin^2\frac{\dt\omega_n}{2}\right)\nn
&&+\frac{2\dt^2\omega_0^2}{mT}\sum_{n=1}^N\frac{1-\sin^2\frac{\dt\omega_n}2}{\frac{4}{\dt^2}\sin^2\frac{\dt\omega_n}2-\omega_0^2+i\epsilon},
\eea
giving
\bea
D_{0++}(-\dt,-\dt)&=&\frac{2\dt}{m\pi}\int_0^\pi d\omega\left(1-\sin^2\frac{\omega}{2}\right)\nn
&&+\frac{2\dt^2\omega_0^2}{m\pi}\int_0^\infty d\omega\frac1{\omega^2-\omega_0^2+i\epsilon}\nn
&=&\frac\dt{m}-i\frac{\dt^2\omega_0}m
\eea
for small $\dt$. We can finally assemble the CTP Green function \eq{ctpgfnt},
\be\label{gfctinft}
\hD(t,t')=-\frac{i}{2m\omega_0}\begin{pmatrix}e^{-i\omega_0|t-t'|}&e^{i\omega_0(t-t')}\cr e^{-i\omega_0(t-t')}&-e^{i\omega_0|t-t'|}\end{pmatrix}.
\ee

\subsection{Quantum open harmonic oscillator}\label{qgrfncta}
The propagator is defined by
\bea
i\hbar D_{\sigma,\sigma'}(t,t')&=&\Tr[\bar T[x_\sigma(t)x_{\sigma'}(t')]\rho_i]\nn
&=&-i\hbar \fdd{W[\hj]}{j_\sigma(t)}{j_{\sigma'}(t')}
\eea
by using the ground state as initial state, $\rho_i=|0\ra\la0|$. One defines the CTP time ordering $\bar T$ according to the first line of eq. \eq{genfunct}: $\bar T$ places the variables $\sigma=-$ in anti-chronological order $T^*[A(t_A)B(t_B)]=\Theta(t_B-t_A)A(t_A)B(t_B)+\Theta(t_A-t_B)B(t_B)A(t_A)$ left of the variables $\sigma=+$ which are brought into chronological order, $T[A(t_A)B(t_B)]=\Theta(t_A-t_B)A(t_A)B(t_B)+\Theta(t_B-t_A)B(t_B)A(t_A)$,
\be
i\hbar\begin{pmatrix}D_{++}(t,t')&D_{+-}(t,t')\cr D_{-+}(t,t')&D_{--}(t,t')\end{pmatrix}=\begin{pmatrix}\la T[x(t)x(t')]\ra&\la x(t')x(t)\ra\cr\la x(t)x(t')\ra&\la T[x(t)x(t')]\ra^*\end{pmatrix}
\ee

The destruction operator $a=(m\omega_0 x+ip)/\sqrt{2m\omega_0}$ acquires the time dependence $a(t)=a(0)e^{-it\omega_0t}$ in the Heisenberg representation. The coordinate operator $x=(a+a^\dagger)/\sqrt{2m\omega_0}$ and the canonical commutation relation $[a,a^\dagger]=1$ leads to the expression \eq{gfctinft} in a well known  manner for the ground state as initial state, in agreement with Ehrenfest's theorem which is exact for harmonic dynamics.

\end{document}